\documentclass{moriond}

\usepackage{graphicx}
\usepackage{braket}
\usepackage[usenames]{color}
\usepackage{latexsym}
\usepackage{bm}
\usepackage{mathrsfs}
\usepackage[caption=false]{subfig}
\usepackage{amsmath} \DeclareMathAlphabet{\mathpzc}{OT1}{pzc}{m}{it}
\usepackage{braket,slashed,bm}

\DeclareMathAlphabet{\mathpzc}{OT1}{pzc}{m}{it}
\newcommand{\Lagr}{\mathcal{L}}

\def\be{\begin{equation}}
\def\ee{\end{equation}}
\def\bea{\begin{eqnarray}}
\def\eea{\end{eqnarray}}

\newcommand{\Photo}{\includegraphics[height=35mm]{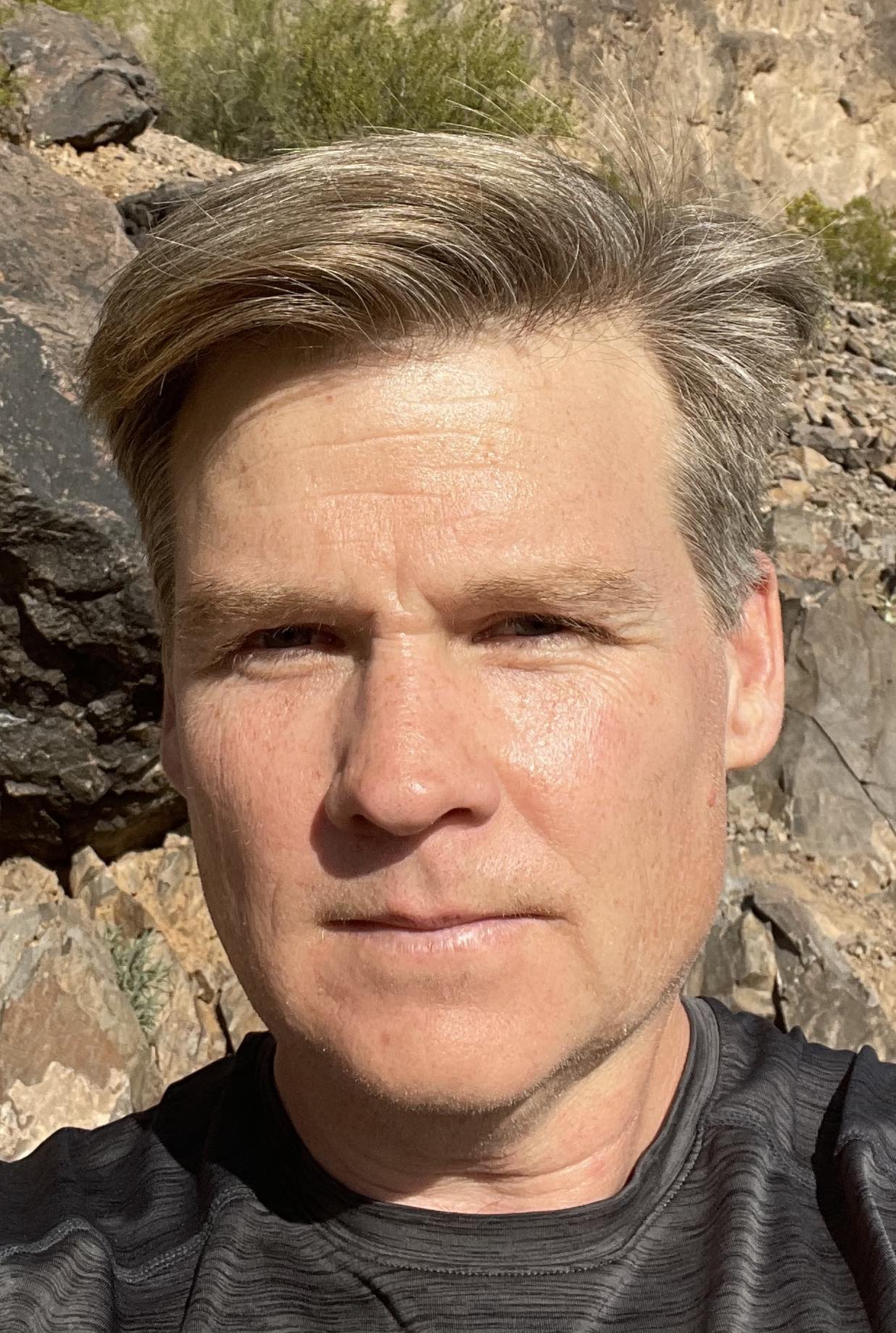}}

\begin{document}
\title{The geometric SMEFT}
\author{Michael Trott }

\address{Niels Bohr Institute, University of Copenhagen,\\
Blegdamsvej 17, DK-2100, Copenhagen, Denmark}
\hspace{0.5cm}

\maketitle\abstracts{Effective field theories, like the Standard Model Effective Field Theory (SMEFT),
are defined by a chosen field content and a set of symmetries, up to a cut off scale $\Lambda$.
Usually, in order to perform calculations, gauge independent
field re-definitions consistent with the symmetries of the theory are then used to redefine
the fields. This procedure results in a fixed (non-redundant) operator basis, that is not itself field re-definition invariant.
Recently, an alternative approach of identifying and calculating with field space
geometry has been developed.
Field redefinition invariants, characterising field space geometry,
appear in observables in amplitude perturbations, and have an expansion in terms of local operators.
In the case of the SMEFT, calculating via the geometric approach is known as the geoSMEFT.
This approach makes it much easier to calculate at high orders in $1/\Lambda$ in the SMEFT,
and can directly result in a complete characterisation of an amplitude perturbation in the $1/\Lambda$
expansion. Using the geoSMEFT, several consistent
and complete $\mathcal{O}(1/\Lambda^4)$ results are now known.
We define the geoSMEFT and demonstrate its use in some examples.}

\section{Introduction}

At this time, there is no statistically significant evidence in particle physics
experiments pointing to an explicit new state to add to Standard Model (SM).
Simultaneously, several interesting hints of deviations from the SM expectation in the global
pattern of experimental results exist. This argues that a mass gap is present
between any putative scale of new physics $\sim \Lambda$ and the electroweak scale $v \simeq 246 \, {\rm GeV}$;
possibly leading to deviations in the predicted SM pattern for some experimental results.
Such a mass gap, if limited to $v/\Lambda < 1/4 \, \pi$ can also be consistent with expectations of UV physics motivated by
naturalness concerns for the Higgs mass ($m_h$).  Broad classes of new physics scenarios
consistent with this assumption can be studied efficiently using
Effective Field Theory (EFT) methods to analyse data sets gathered at energies $\sqrt{s} \sim v << \Lambda$.

It is of interest to examine current, and possible future deviations in measurements, in a consistent theoretical EFT framework,
that also incorporates lower energy experimental data gathered on the $Z, \, W^\pm$ and $h$ particle phase space poles.
This theoretical framework has come to be known as the linear Standard Model
Effective Field Theory (SMEFT). Here ``linear SMEFT" refers to the assumption that the
particle spectrum contains a $\rm SU(2)$ scalar doublet $H$, before the Higgs takes on a vacuum expectation value. Developing
predictions for particle physics experiments in the linear SMEFT to $\mathcal{O}(1/\Lambda^2)$
is essentially a solved problem\cite{Alonso:2013hga,Brivio:2017vri},
and fully automated when using the SMEFTsim package\cite{Brivio:2017btx,Brivio:2020onw}.
However, the SMEFT is an expansion in $1/\Lambda$. When interpreting the data
at $\mathcal{O}(1/\Lambda^2)$ in the SMEFT, the sub-leading $\mathcal{O}(1/\Lambda^4)$ terms in the expansion
exist and cannot be ignored if the relative suppression between these corrections
is not numerically negligible. This is likely to be the case if deviations from the SM are found.
As such, the effects of these sub-leading terms should be considered, not naively ignored.
In addition, loop corrections in the SMEFT also exist, which also need consideration, and at times
should be incorporated into global SMEFT fits\cite{Brivio:2019ius,Ellis:2020unq,Almeida:2021asy,Ethier:2021bye}.

The development of SMEFT loop calculations has advanced steadily for years,
since the foundation of developing these corrections was laid down by the systematic dimension six
(gauge independent) renormalization results\cite{Alonso:2013hga,Jenkins:2013zja,Jenkins:2013wua}.
In recent years, a dimension eight operator basis has also been defined and published\cite{Murphy:2020rsh}.
However, despite initial studies and partial results\cite{Hays:2018zze}, systematically advancing the predictions
of the SMEFT to dimension eight was a daunting prospect. Such theoretical calculations
are of value to have an informed understanding of neglected dimension eight terms
in SMEFT studies at $\mathcal{O}(1/\Lambda^2)$ to avoid over interpreting the results of global SMEFT fits.
This point has long been made in the
literature\cite{Passarino:2012cb,David:2013gaa,Passarino:2016pzb,Berthier:2015oma}
when considering the SMEFT as a model independent (bottom up) EFT. To define such error {\it estimates} in a bottom up fashion within the EFT
here we focus on how to develop theoretical calculations consistently
to $\mathcal{O}(1/\Lambda^4)$ rapidly, completely and efficiently.\footnote{Note that an alternative approach of assuming a
specific UV completion, or a set of UV completions (i.e. insisting on a top down understanding), to fix matching patterns is doomed to fail in characterising
the full space of the SMEFT as a bottom up EFT. As then one is assuming {\it a different theory} then the SMEFT
to define the effect of sub-leading terms. Asserting that only a top down EFT approach is possible for error estimates
in EFT studies\cite{Brivio:2022pyi}, or EFT in general, would be as absurd as asserting only a bottom up EFT approach is
possible for such studies.
The approach pursued for an error {\it estimate} is always a freely chosen convention.
Here, we report and summarise results in some recent published literature\cite{Trott:2021vqa}
developed (in part) to enable robust error estimates in a bottom up approach to the SMEFT.}

The key to enabling the systematic advance to $\mathcal{O}(1/\Lambda^4)$ was to reconsider a basic feature in the SMEFT.
Field redefinitions are used to define an operator basis in an effective field theory. But the
resulting operator basis is {\it not} field redefinition invariant. This is well known and not an intrinsic problem
at any mass dimension. An interesting question is: ``Is there a way
to calculate that is more field redefinition invariant?" It is well known that operator bases
are unphysical, in that they combine up in a contribution to a specific observable in a manner
that a consistent global pattern of deviations are present -- irrespective of the operator basis chosen.
This implies that combinations of operators
can be considered more ``physical", than an individual operator. The immediate follow up question then becomes:
- ``Are particular combinations of operators more useful to consider?"
And if so - what defines such combinations of operators systematically?

A useful set of intermediate geometric quantities that the operators combine into in the SMEFT are now known.
The key point is that abstract interaction field spaces
are present in the EFT and define tensors - metrics and connections on such spaces. These tensors are useful as they then
project geometric invariants of these spaces onto amplitudes perturbations
in the SMEFT. It is these geometric invariants the the operators combine up into in amplitude perturbations
(as observables are field redefinition invariant).
When considering low n-point interactions critical for phenomenology, the number of geometric quantities
that so project is limited and tractable to identify. Defining and exploiting these geometric quantities leads to
the geometric formulation of the SMEFT (geoSMEFT).

\subsection{geoSMEFT}\label{subsec:geoSMEFT}

Building on many
key steps in the previous
literature\cite{Burgess:2010zq,Alonso:2015fsp,Alonso:2016btr,Alonso:2016oah,Brivio:2017vri,Helset:2018fgq,Corbett:2019cwl}
The geoSMEFT\cite{Helset:2020yio} is constructed in the following fashion.
For the sake of constructing CP even two, three point functions,
a set of field space metrics and connections are defined as follows. The scalar potential is
\bea
V(\phi) =  \left.-\Lagr_{\rm SMEFT}\right|_{\Lagr(\alpha,\beta\cdots\rightarrow 0)}.
\eea
The field space metric for the scalar field bilinear, dependent on the SM field coordinates, is defined via
\bea\label{hijdefn}
h_{IJ}(\phi) = \left.\frac{g^{\mu \nu}}{d} \, \frac{\delta^2 \mathcal{L}_{\rm SMEFT}}{\delta(D_\mu \phi)^I \, \delta (D_\nu \phi)^J} \right|_{\mathcal{L}(\alpha,\beta \cdots) \rightarrow 0}.
\eea
The notation $\mathcal{L}(\alpha,\beta \cdots)$ corresponds to non-trivial Lorentz-index-carrying
Lagrangian terms and spin connections, i.e. $\{\mathcal{W}_{\mu \nu}^A, (D^\mu \Phi)^K, \bar{\psi} \sigma^\mu \psi, \bar{\psi} \psi \cdots \}$.
Here $d=4$ is the number of space time dimensions and $g^{\mu\nu}$ is the usual Minkowskian space time metric.
Note that this definition reduces the connection $h_{IJ}$ to a function
of $\rm SU(2)_L \times U(1)_Y$ generators, scalar fields coordinates $\phi_i$ and $\bar{v}_T$.
Here $\sqrt{2 \langle H^\dagger H \rangle_{SM}} \equiv \bar{v}_T$ is the vev, including the tower of higher
order corrections in the SMEFT.
The CP even gauge field scalar manifolds, for the ${\rm SU(2)_L \times U(1)_Y}$ fields interacting with the scalar fields,
give
 \bea
g_{AB}(\phi)
=  \left. \frac{-2 \, g^{\mu \nu} \, g^{\sigma \rho}}{d^2}
 \, \frac{\delta^2 \mathcal{L}_{\rm SMEFT}}{\delta \mathcal{W}^A_{\mu \sigma} \, \delta \mathcal{W}^B_{\nu \rho}}
 \, \right|_{\mathcal{L}(\alpha,\beta \cdots) \rightarrow 0,{\textrm{CP-even}}},
\eea
and (here $ \mathpzc{A}, \mathpzc{B}$ run over $1 \cdots 8$)
\bea
k_{\mathpzc{AB}}(\phi) =  \left.\frac{-2 \, g^{\mu \nu} \, g^{\sigma \rho}}{d^2} \, \frac{\delta^2 \mathcal{L}_{\rm SMEFT}}{\delta G^{\mathpzc{A}}_{\mu \sigma} \, \delta G^{\mathpzc{B}}_{\nu \rho}} \,\right|_{\mathcal{L}(\alpha,\beta \cdots) \rightarrow 0,{\textrm{CP-even}}}.
\eea
We also have
\bea
k_{IJ}^A(\phi) =\left. \frac{g^{\mu\rho}g^{\nu\sigma}}{2d^2} \frac{\delta^3 \mathcal{L}_{\rm SMEFT}}{\delta (D_\mu \phi)^I \delta (D_\nu \phi)^J \mathcal{W}_{\rho\sigma}^A}\right|_{\mathcal{L}(\alpha,\beta \cdots) \rightarrow 0}
\eea
and
\bea
f_{ABC}(\phi) = \left.\frac{g^{\nu\rho}g^{\sigma\alpha}g^{\beta\mu}}{3!d^3} \frac{\delta^3\mathcal{L}_{\rm SMEFT}}{\delta \mathcal{W}_{\mu\nu}^A \delta \mathcal{W}_{\rho\sigma}^B \mathcal{W}_{\alpha\beta}^C}\right|_{\mathcal{L}(\alpha,\beta\cdots)\rightarrow 0,\textrm{CP-even}},
\nonumber \\
k_{\mathpzc{ABC}}(\phi) = \left.\frac{g^{\nu\rho}g^{\sigma\alpha}g^{\beta\mu}}{3!d^3} \frac{\delta^3\mathcal{L}_{\rm SMEFT}}{\delta G_{\mu\nu}^{\mathpzc{A}} \delta G_{\rho\sigma}^{\mathpzc{B}} G_{\alpha\beta}^{\mathpzc{C}}}\right|_{\mathcal{L}(\alpha,\beta\cdots)\rightarrow 0,\textrm{CP-even}}.
\eea

We also define the fermionic connections
\bea \label{eq:DefYukawa}
Y^{\psi_1}_{pr}(\phi_I) =  \left. \frac{\delta \mathcal{L}_{\rm SMEFT}}{\delta (\bar{\psi}^I_{2,p} \psi_{1,r})} \right|_{\mathcal{L}(\alpha,\beta \cdots) \rightarrow 0},
\quad
L_{J,A}^{\psi,pr} = \left. \frac{\delta^2 \mathcal{L}_{\rm SMEFT}}{\delta (D^\mu \phi)^J \delta (\bar{\psi}_{p} \gamma_\mu \tau_A \psi_r)}  \right|_{\mathcal{L}(\alpha,\beta \cdots) \rightarrow 0},
\eea
and
\bea
d_A^{\psi_1, pr}(\phi_I) = \left. \frac{\delta^2 \mathcal{L}_{\rm SMEFT}}{\delta (\bar{\psi}_{2,p}^I \sigma_{\mu \nu} \psi_{1,r}) \delta \mathcal{W}_{\mu \nu}^A} \right|_{\mathcal{L}(\alpha,\beta \cdots) \rightarrow 0}.
\eea

The explicit form of these field space metrics and connects are given in the core geoSMEFT paper\cite{Helset:2020yio}
{\it to all orders in $\bar{v}_T/\Lambda$.} The remarkable compactness of the all orders form of these objects
in a key expansion parameter for precision SMEFT phenomenology,
is due to the fact that they are functions of arbitrary numbers of Higgs field emissions,
and symmetry generators acting on these fields. The completeness
relations of the $\rm SU(2)\times U(1)$ algebras reduce the complexity of these objects to a compact form
whose complexity usually saturates around mass dimension eight -- when considering operator expansions of the metrics
and tensors. This can also be seen in Table~\ref{tab:table1}.

\begin{table}
\caption[]{
\label{tab:table1}Counting of operators contributing to two- and three-point functions from Hilbert series\cite{Lehman:2015via,Henning:2015alf}.\\
$N_f$ corresponds to the number of fermion generations.}
\begin{center}
\begin{tabular}{|c|c|c|c|c|c|}
\hline
\multicolumn{1}{|c}{} &
\multicolumn{5}{|c|}{\textrm Mass Dimension} \\
\hline
\multicolumn{1}{|c|}{\textrm Field space connection}
 & 6 & 8 & 10 & 12 & 14 \\
 \hline
 $h_{IJ}(\phi)(D_\mu\phi)^I (D^\mu \phi)^J$ & 2 & 2 & 2 & 2 & 2 \\
 $g_{AB}(\phi)\mathcal{W}_{\mu\nu}^A \mathcal{W}^{B,\mu\nu}$ & 3 & 4 & 4 & 4 & 4 \\
 $k_{IJA}(\phi)(D^\mu \phi)^I(D^\nu\phi)^J \mathcal{W}_{\mu\nu}^A$ & 0 & 3 & 4 & 4 & 4 \\
 $f_{ABC}(\phi) \mathcal{W}_{\mu\nu}^A \mathcal{W}^{B,\nu\rho} \mathcal{W}_{\rho\mu}^C $ & 1 & 2 & 2 & 2 & 2 \\
 $Y^{u}_{pr}(\phi) \bar{Q} u +$ {\textrm h.c.} & $2 \, N_f^2$ & $2 \, N_f^2$ & $2 \, N_f^2$ & $2 \, N_f^2$ & $2 \, N_f^2$ \\
 $Y^{d}_{pr}(\phi) \bar{Q} d +$ {\textrm h.c.} & $2 \, N_f^2$ & $2 \, N_f^2$ & $2 \, N_f^2$ & $2 \, N_f^2$ & $2 \, N_f^2$ \\
 $Y^{e}_{pr}(\phi) \bar{L} e +$ {\textrm h.c.} & $2 \, N_f^2$ & $2 \, N_f^2$ & $2 \, N_f^2$ & $2 \, N_f^2$ & $2 \, N_f^2$ \\
 $d^{e}_{pr}(\phi) \bar{L} \sigma_{\mu \nu} e \mathcal{W}^{\mu \nu}_A +$ {\textrm h.c.} & $4 \, N_f^2$ & $6 \, N_f^2$ & $6 \, N_f^2$ & $6 \, N_f^2$ & $6 \, N_f^2$ \\
$d^{u}_{pr}(\phi) \bar{Q} \sigma_{\mu \nu} u \mathcal{W}^{\mu \nu}_A +$ {\textrm h.c.} & $4 \, N_f^2$ & $6 \, N_f^2$ & $6 \, N_f^2$ & $6 \, N_f^2$ & $6 \, N_f^2$ \\
$d^{d}_{pr}(\phi) \bar{Q} \sigma_{\mu \nu} d \mathcal{W}^{\mu \nu}_A +$ {\textrm h.c.} & $4 \, N_f^2$ & $6 \, N_f^2$ & $6 \, N_f^2$ & $6 \, N_f^2$ & $6 \, N_f^2$ \\
$L^{\psi_R}_{pr,A}(\phi) (D^\mu \phi)^J  (\bar{\psi}_{p,R} \gamma_\mu \tau_A \psi_{r,R}) $  & $N_f^2$ & $N_f^2$ & $N_f^2$ & $N_f^2$ & $N_f^2$ \\
$L^{\psi_L}_{pr,A}(\phi) (D^\mu \phi)^J  (\bar{\psi}_{p,L} \gamma_\mu \tau_A \psi_{r,L}) $  & $2 \, N_f^2$ & $4 \, N_f^2$ & $4 \, N_f^2$ & $4 \, N_f^2$ & $4 \, N_f^2$
\\ \hline
 \end{tabular}
 \end{center}
 \end{table}

\subsection{Mass eigenstate transformations at all orders in $\bar{v}_T/\Lambda$}
The utility of understanding the SMEFT through the field space geometric quantities present
is multi-fold. For example,
one can immediately confirm that the transformation at all orders in $\bar{v}_T/\Lambda$
between weak and mass eigenstates can be compactly expressed as
\bea\label{massweak}
 \alpha^A = \mathcal{U}^{A}_C \, \beta^C,  \quad  \mathcal{W}^{A,\mu} =  \mathcal{U}^{A}_C \mathcal{A}^{C,\mu},   \quad  \phi^J = \mathcal{V}^{J}_K \, \Phi^K,
\eea
where in the SM limit $\alpha^A = \{g_2 \, g_2, g_2, g_1 \}$ and $\mathcal{W}^A = \{W_1,W_2,W_3,B\}$
and
\bea
\beta^C &=& \left\{\frac{g_2 \,(1-i)}{\sqrt{2}}, \frac{g_2 \,(1+i)}{\sqrt{2}}, \sqrt{g_1^2+ g_2^2}(c_{\bar{\theta}}^2 - s_{\bar{\theta}}^2), \frac{2 \, g_1 \, g_2}{\sqrt{g_1^2+ g_2^2}} \right\}, \\
\mathcal{A}^{C} &=& \left(\mathcal{W}^+,\mathcal{W}^-, \mathcal{Z}, \mathcal{A}\right).
\eea
Here $ \phi^J = \{\phi_1,\phi_2,\phi_3,\phi_4 \}$, $\Phi^K = \{\Phi^-,\Phi^+,\chi, h \}$ for the scalar fields
with normalisation
\bea
H(\phi_I) = \frac{1}{\sqrt{2}}
\begin{bmatrix} \phi_2+i\phi_1 \\ \phi_4 - i\phi_3\end{bmatrix}, \quad \tilde{H}(\phi_I) = \frac{1}{\sqrt{2}} \begin{bmatrix} \phi_4 + i\phi_3 \\ - \phi_2+i\phi_1\end{bmatrix}.
\eea
Note $\phi_4$ is expanded around the vacuum expectation value with the replacement $\phi_4 \rightarrow \phi_4 + \bar{v}_T$
and
\begin{align*}
	U_{BC} &= \begin{bmatrix}
		\frac{1}{\sqrt{2}} & \frac{1}{\sqrt{2}} & 0 & 0 \\
		\frac{i}{\sqrt{2}} & \frac{-i}{\sqrt{2}} & 0 & 0 \\
		0 & 0 & c_{\overline{\theta}} & s_{\overline{\theta}} \\
		0 & 0 & -s_{\overline{\theta}} & c_{\overline{\theta}}
	\end{bmatrix},& \quad
	V_{JK} &= \begin{bmatrix}
		\frac{-i}{\sqrt{2}} & \frac{i}{\sqrt{2}} & 0 & 0 \\
		\frac{1}{\sqrt{2}} & \frac{1}{\sqrt{2}} & 0 & 0 \\
	0 & 0 & -1 & 0 \\
	0 & 0 & 0 & 1
	\end{bmatrix}.
\end{align*}
We are using a short-hand notation  for the transformation matrices
that lead to the canonically normalized mass eigenstate fields
 \begin{align*}
 \mathcal{U}^{A}_C &= \sqrt{g}^{AB} U_{BC}, & \quad  \mathcal{V}^{I}_K &= \sqrt{h}^{IJ} V_{JK}.
 \end{align*}
 Here $\sqrt{g}^{AB}$ and $\sqrt{h}^{IJ}$ are square-root metrics, which are understood to be matrix square roots
 of the expectation value --  $\langle \rangle$ -- of the field space connections for the bilinear terms in the SMEFT.
Also note $ \sqrt{h}^{IJ} \sqrt{h}_{JK}\equiv \delta^I_K$ and $ \sqrt{g}^{AB} \sqrt{g}_{BC}\equiv \delta^A_C$.
The rotation angles $c_{\overline{\theta}}, s_{\overline{\theta}}$ are functions of $\alpha_A$ and $\langle g^{AB} \rangle$
and are defined geometrically in the core geoSMEFT paper\cite{Helset:2020yio}.

\subsection{The $h \mathcal{A}\mathcal{A}$ amplitude perturbation}
One can also express amplitude perturbations at all orders in the $\bar{v}_T/\Lambda$ expansion of the theory.
For example, the effective coupling of {h-$\gamma$-$\gamma$},
including the tower of $\bar{v}_T^2/\Lambda^2$ corrections, is given by
\bea
\langle h| \mathcal{A}(p_1) \mathcal{A}(p_2) \rangle = -\langle h A^{\mu\nu} A_{\mu \nu} \rangle \frac{\sqrt{h}^{44}}{4} \left[
	\langle \frac{\delta g_{33}(\phi)}{\delta \phi_4}\rangle \frac{\overline{e}^2}{g_2^2}  +
 2\langle \frac{\delta g_{34}(\phi)}{\delta \phi_4}\rangle \frac{\overline{e}^2}{g_1 g_2}  +
  \langle \frac{\delta g_{44}(\phi)}{\delta \phi_4}\rangle \frac{\overline{e}^2}{g_1^2}
	\right], \nonumber
\eea
where
$\bar{e} = g_2\left(s_{\bar{\theta}} \sqrt{g}^{33} + c_{\bar{\theta}} \sqrt{g}^{34}  \right) = g_1\left(c_{\bar{\theta}} \sqrt{g}^{44} + s_{\bar{\theta}} \sqrt{g}^{34}  \right).$

\subsection{$\mathcal{W},\mathcal{Z}$ couplings to $\bar{\psi} \psi$}
The mass eigenstate coupling of the $\mathcal{Z}$ and $\mathcal{W}$ to $\bar{\psi} \psi$ are obtained by summing over more than one field space connection.
For couplings to fermion fields of the same chirality, the sum is over $L_{J,A}^{\psi,pr}$ and the modified $\bar{\psi} i \slashed{D} \psi$, that includes the tower of SMEFT corrections in
$\mathcal{U}^A_C$.
A compact expression for the mass eigenstate connection is
\bea
- \mathcal{A}^{A,\mu} (\bar{\psi}_{p} \gamma_\mu \bar{\tau}_A \psi_r) \delta_{pr} + \mathcal{A}^{C,\mu}
(\bar{\psi}_{p} \gamma_\mu \sigma_A\psi_r) \langle L_{I,A}^{\psi,pr}\rangle (- {\bm \gamma}^{I}_{C,4}) \bar{v}_T,
\eea
where the fermions are in the weak eigenstate basis. Rotating the fermions to the mass eigenstate basis is straightforward, where the $V_{\rm CKM}$ and $U_{\rm PMNS}$
matrices are introduced as usual\cite{Talbert:2021iqn}.
Expanding out to make the couplings explicit, the Lagrangian effective couplings for $\{\mathcal{Z},\mathcal{A},\mathcal{W}^\pm\}$ are
\bea
\langle \mathcal{Z} | \bar{\psi}_{\substack{p}} \psi_{\substack{r}}\rangle &=&
\frac{\bar{g}_Z}{2} \, \bar{\psi}_{\substack{p}} \, \slashed{\epsilon}_{\mathcal{Z}} \, \left[(2 s_{\theta_Z}^2  Q_\psi  - \sigma_3)\delta_{pr}
	+\sigma_3\bar{v}_T \langle L_{3,3}^{\psi,pr}\rangle+ \bar{v}_T \langle L_{3,4}^{\psi,pr} \rangle
 \right] \, \psi_{\substack{r}}, \\
\langle \mathcal{A} | \bar{\psi}_{\substack{p}} \psi_{\substack{r}} \rangle &=&
- \bar{e} \, \bar{\psi}_{\substack{p}} \, \slashed{\epsilon}_{\mathcal{A}} \, Q_\psi \, \delta_{pr}\, \psi_{\substack{r}}, \\
\langle \mathcal{W}_{\pm} | \bar{\psi}_{\substack{p}} \psi_{\substack{r}} \rangle &=& - \frac{\bar{g}_2  }{\sqrt{2}}
\bar{\psi}_{\substack{p}}  (\slashed{\epsilon}_{\mathcal{W}^{\pm}}) \, T^{\pm}
\left[\delta_{pr}-\bar{v}_T \langle L^{\psi,pr}_{1,1}\rangle   \pm  i \bar{v}_T
\langle L^{\psi,pr}_{1,2} \rangle \right]
\, \psi_{\substack{r}}.
\eea
The last expressions simplify due to $\rm SU(2)_L$ gauge invariance.
Similarly the SMEFT has the right-handed $W^\pm$ couplings to (weak eigenstate) quark fields.
\begin{align*}
\langle \mathcal{W}_{+}^\mu | \bar{u}_{\substack{p}} d_{\substack{r}} \rangle &=   \bar{v}_T \langle L^{ud,pr}_1 \rangle \frac{\bar{g}_2}{\sqrt{2}} \bar{u}_p \,\slashed{\epsilon}_{\mathcal{W}^+} d_r, & \langle \mathcal{W}_{-}^\mu | \bar{d}_{\substack{r}} u_{\substack{p}} \rangle &=   \bar{v}_T \langle L^{ud,pr}_1 \rangle \frac{\bar{g}_2}{\sqrt{2}} \bar{d}_r \,\slashed{\epsilon}_{\mathcal{W}^-} u_p.
\end{align*}

Using these expressions, EWPD was analysed for the first time
to dimension eight\cite{Corbett:2021eux}.
Dirac masses, mixings, and CP-violation parameter\cite{Talbert:2021iqn} were directly also derived in the geoSMEFT.
The geoSMEFT has also been used to calculate consistent expressions\cite{Corbett:2021cil,Martin:2021cvs} for
$\sigma(\mathcal{G} \,\mathcal{G}\rightarrow h)$ and $\Gamma(h \rightarrow \mathcal{G} \,\mathcal{G})$
to $\mathcal{O}(1/\Lambda^4)$ while simultaneously developing the corresponding amplitude results to
$\mathcal{O}(1/16 \pi^2 \Lambda^2)$ using the background field method\cite{Helset:2018fgq}
and it was noted that these two sub-leading terms in the SMEFT are not independent.

It is interesting
that the loop and operator
mass dimension series expansions are
cross correlated in order to maintain gauge invariance in the SMEFT, i.e. satisfy the corresponding
Ward identity relations between n-point functions\cite{Corbett:2019cwl} order by order in the
$1/\Lambda$ power
counting in the theory.
This fact is explicitly demonstrated in some examples at one loop in the literature\cite{Corbett:2020ymv,Corbett:2020bqv}.
It should also be intuitive that these expansions are not independent in general, as gauge fixing
acts on fields that themselves are redefined order by order in the operator expansion of the theory
when the Higgs takes on a vacuum expectation value. For this basic reason, strong general claims on gauge independence of results in the SMEFT
at $\mathcal{O}(1/\Lambda^{4})$ between n-point functions, simply due to squaring
gauge (parameter) independent results at $\mathcal{O}(1/\Lambda^{2})$ clearly require detailed mathematical demonstrations,
and cannot be established only by assertion. The interested reader is encouraged
to examine some recent debate\cite{Brivio:2022pyi,Trott:2021vqa,Corbett:2019cwl,Corbett:2020ymv,Corbett:2020bqv}
on this issue present in the literature, from dramatically different perspectives.

\section*{Conclusions}
The geoSMEFT is a consequence of the field redefinition invariance defining the SMEFT.
It is a manifestly useful construction, allowing all order results in the $\bar{v}_T/\Lambda$ expansion of the theory
to be directly and immediately developed. In the literature, these theoretical
techniques have already pushed calculations in the SMEFT to unprecedented theoretical precision in multiple processes,
and many interesting formal questions have been investigated using field space geometry in recent
months\cite{Cheung:2022vnd,Cheung:2021yog,Cohen:2022uuw}. Using field space geometries
to ones advantage can be the key to efficient sub-leading order calculation in the SMEFT, and other field theories defined
by field re-definitions. These results critically inform efforts to perform consistent
fits to global data sets including a full set of $\mathcal{O}(1/\Lambda^2)$ corrections, by defining
the corresponding SMEFT corrections to the SM to $\mathcal{O}(1/\Lambda^4)$ rapidly, completely and efficiently.
This allows us to better understand the effect of neglected higher order terms when studying the global data
set in the SMEFT. Fitting and constraining parameters introduced as corrections to the SM at $\mathcal{O}(1/\Lambda^2)$
should be maximally informed by these results for robust bottom up SMEFT studies, which are expected to be a core legacy
of the Large Hadron Collider physics program.

\section*{Acknowledgments}

The author thanks the Villum Fonden project number 00010102. The Danish National Research Foundation (DNRF91) through the Discovery center, and DFF for support.

\section*{References}
\bibliography{bibliography.bib}

\end{document}